\documentclass[sigconf,nonacm]{acmart}
\AtBeginDocument{%
  }

\setcopyright{acmlicensed}
\copyrightyear{2018}
\acmYear{2018}
\acmDOI{XXXXXXX.XXXXXXX}
\acmConference[Conference acronym 'XX]{Make sure to enter the correct
  conference title from your rights confirmation email}{June 03--05,
  2018}{Woodstock, NY}
\acmISBN{978-1-4503-XXXX-X/2018/06}

\usepackage{amsmath,amsfonts}
\usepackage{algorithmic}
\usepackage{textcomp}
\usepackage{xcolor}
\usepackage{amsthm}
\usepackage{graphicx}
\usepackage{booktabs}
\usepackage{array}
\usepackage{bm}
\usepackage{multirow}
\usepackage{textcomp}
\usepackage{multicol} 
\usepackage{pifont}
\usepackage{algorithm}
\usepackage{pifont}
\usepackage{threeparttable}
\usepackage{algorithmic}
\usepackage{breakurl}
\usepackage{hyperref}
\hypersetup{
    colorlinks=false, % 禁用链接颜色
    pdfborder={0 0 0} % 设置边框为无
}
\begin{document}

\title[Dynamic Airspace Management for UAVs in Evolving Urban Environments]{Dynamic Airspace Management for UAVs in
Evolving Urban Environments: Collaborative Coordination and Human Safety}
% 第一块：交大作者 1（共同一作）
% =================================================================
\author{Lin Sun}
\authornote{These authors contributed equally to this research.} % 自动生成脚注1：共同一作
\affiliation{%
  \institution{Shanghai Jiao Tong University}
  \city{Shanghai}
  \country{China}
}
\email{sun_lin@sjtu.edu.cn}

% =================================================================
% 第二块：交大作者 2（共同一作）
% =================================================================
\author{Yuhang Wang}
\authornotemark[1] % 复用脚注1（共同一作）
\affiliation{%
  \institution{Shanghai Jiao Tong University}
  \city{Shanghai}
  \country{China}
}
\email{lingbo_2022@sjtu.edu.cn}

% =================================================================
% 第三块：交大作者 3（共同一作）
% =================================================================
\author{Fan Meng Hong}
\authornotemark[1] % 复用脚注1（共同一作）
\affiliation{%
  \institution{Shanghai Jiao Tong University}
  \city{Shanghai}
  \country{China}
}
\email{stevenhong@sjtu.edu.cn}

% =================================================================
% 第四块：交大作者 4（通信作者）
% =================================================================
\author{Haopeng Chen}
\authornote{Corresponding author.} % 自动生成脚注2：通信作者
\affiliation{%
  \institution{Shanghai Jiao Tong University}
  \city{Shanghai}
  \country{China}
}
\email{chen-hp@sjtu.edu.cn}

% 第五个：企业作者 1（Yan Jiao）
% =================================================================
\author{Yan Jiao}
\affiliation{%
  \institution{Shanghai Shapere Information Technology Co.,Ltd.}
  \city{Shanghai}
  \country{China}
}
\email{jiaoyan2026@126.com}

% =================================================================
% 第六个：企业作者 2（Yongming Xu）
% =================================================================
\author{Yongming Xu}
\affiliation{%
  \institution{Shanghai Shapere Information Technology Co.,Ltd.}
  \city{Shanghai}
  \country{China}
}
\email{xuym2026@126.com}

\renewcommand{\shortauthors}{Lin et al.}

\begin{abstract}
 The low-altitude economy is an emerging industry with significant development potential, in which the safety of unmanned aerial vehicle (UAV) operations is a critical challenge. Particularly within complex urban topographies and human-populated environments, UAV airspace management must prioritize collision avoidance and human safety. We propose Pharos, a collaborative multi-UAV airspace management system. Pharos lies between the distributed local perception paradigm and the centralized fine-grained control paradigm. Pharos coordinates the safe parallel execution of UAVs in shared airspace while innovatively accounting for the impact of human fear. Pharos is implemented using the MAPPO algorithm due to its faster convergence and higher rewards than other typical MARL algorithms (HAPPO and HATRPO). To evaluate Pharos, we developed a 3D simulation system using real urban data. Visualization results demonstrate its effective airspace coordination capability. Regarding performance verification, Pharos reduced human fear by 52.72\% compared to the benchmark Ipopt. Moreover, we designed spatial entropy as a system evaluation metric to quantify space utilization, which improved performance by 70.82\% and 2.03\% compared to the benchmarks Ipopt and A-star, respectively. The source code is available at an anonymized repository: \textit{\url{https://github.com/pharos-anonymized/source-code.git}}.
\end{abstract}




%\received{01 May 2026}

\maketitle
\section{Introduction}

%The low-altitude economy has become a key driver of the contemporary economy, spurred by the advent of unmanned aerial vehicles (UAVs) \cite{zhou2025unmanned}. This sector fosters innovative business models and generates new economic growth vectors, with applications including heavy-lift UAVs for long-haul logistics, lightweight UAVs for infrastructure inspections, and medium-sized UAVs for urban last-mile delivery of food and parcels. 

Driven by the rapid evolution of unmanned aerial vehicles (UAVs), the low-altitude economy has emerged as a highly active domain in recent years \cite{zhou2025unmanned}. Diverse UAV platforms are now increasingly integrated into daily operations, ranging from heavy-lift UAVs for long-haul logistics to lightweight UAVs for infrastructure inspection and urban last-mile delivery. However, when multiple UAVs operate in parallel within a shared low-altitude space, their respective private operations are highly likely to conflict. When UAV flight zones overlap densely populated urban areas, the risk of UAV collisions and the threat to human safety they pose will increase significantly \cite{pereira2022improving}. Consequently, establishing an airspace management system for UAVs to ensure safety at low altitudes is a pressing concern. This involves tackling two main challenges: exploring effective multi-UAV collaborative coordination solutions \cite{chen2025reinforcement} and addressing human fears related to low-altitude flights \cite{van2023increasing}.

 \begin{figure}[t]
\centering
\includegraphics[width=0.49\textwidth]{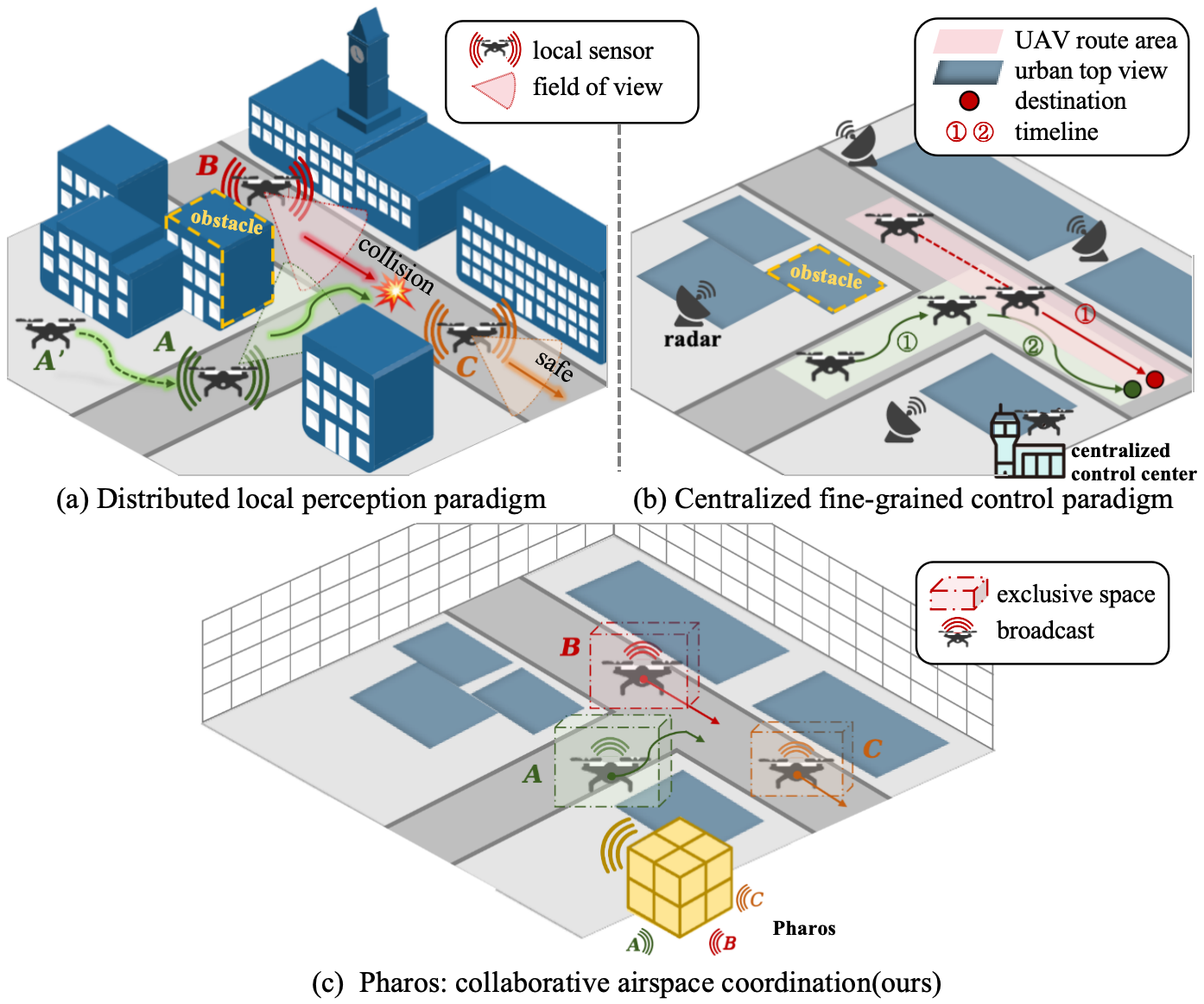} 
\caption{Three airspace management paradigms: (\textit{a}) Distributed local perception paradigm may lead to collisions of $A$,$B$ at blind spots. (\textit{b}) Centralized fine-grained control paradigm ensures safety for $A$ and $B$ via sequential instructions by timeline \ding{172}-\ding{173}. (\textit{c}) Pharos coordinates the exclusive spaces for $A$ and $B$ through coarse-grained global management to enable safe parallel passage.}
\label{fig:intro_overview}
\vspace{-0.5cm}
\end{figure}

Firstly, human safety factors are currently mainly considered in applications involving two-dimensional spaces, such as autonomous driving \cite{zhao2024human} and industrial robotics \cite{merckaert2024real}. Unlike the relatively predictable roadways for autonomous vehicles or the enclosed and controlled environments for industrial robots, UAVs operate in open, three-dimensional environments. In these dynamic, integrated settings where UAVs and humans coexist, addressing human safety issues in UAV airspace management has become increasingly complex. There is an urgent need to explore solutions to fill this gap in the research field.

Secondly, there are currently two main paradigms for UAV collaborative coordination. One is the \textbf{distributed local perception paradigm}, as typically seen in conventional Embodied Intelligence Systems (EAI) \cite{hu2024toward}. This involves equipping UAVs with local sensors, enabling them to operate based on information from their own fields of view. While effective for \textbf{individual autonomy}, the reliance on onboard sensors introduces substantial payload and cost overheads, which can be restrictive for the \textbf{scalable} deployment of \textbf{lightweight} UAVs. Moreover, the local field of view provided by onboard sensors inherently \textit{limits global situational awareness}, potentially exposing UAVs to blind-spot collisions in the presence of obstacles. For instance, high-rise buildings may impede the mutual detection of UAVs $A$ and $B$ at key intersections, as shown in Fig. \ref{fig:intro_overview}(a). The other paradigm is \textbf{centralized fine-grained control paradigm}, such as the traditional Air Traffic Control (ATC) technology \cite{huang2024code}. As in Fig. \ref{fig:intro_overview}(b), the ATC center integrates radar data and UAV telemetry to maintain global situational awareness and directly issues flight commands to UAVs through \textbf{fine-grained trajectory} planning algorithms \cite{meng2024ppswarm,basil2025performance,fu2024red}. While this paradigm enhances the safety of UAVs' collaboration, it faces implementation challenges. These include the need for standardized \textbf{control interfaces} from UAV manufacturers and potential performance bottlenecks at the centralized control center. 

%with resource-intensive perception payloads (e.g., LiDARs and depth cameras)

Therefore, a solution is needed to bridge the two paradigms above by coordinating the safe parallel execution of multiple UAVs. To this end, we propose \textbf{Pharos}, a novel collaborative multi-UAV airspace management paradigm. As shown in Fig. \ref{fig:intro_overview}(c), Pharos achieves this \textbf{coordination} not by directly planning detailed trajectories, but by coordinating \textbf{exclusive spaces} for each UAV, allowing their autonomous operations within the safe spaces. By doing so, \textbf{Pharos} effectively overcomes the limitations of the \textbf{distributed local perception paradigm}. It reduces collision risks arising from \textbf{incomplete} local views and the payload and cost constraints caused by advanced onboard sensors, specifically \textbf{resource-intensive} perception payloads (e.g., LiDARs and depth cameras). Compared to the \textbf{centralized fine-grained control paradigm}, \textbf{Pharos} employs a \textbf{coarse-grained} airspace coordination scheme to alleviate the performance bottlenecks typically caused by \textbf{fine-grained} trajectory planning. Furthermore, by avoiding direct motion control, Pharos does not interfere with the UAVs’ \textbf{underlying control} systems. It provides collaborative guidance relying on a communication interface for \textbf{data exchange}. This eliminates the challenge of requiring UAV manufacturers to open their proprietary \textbf{control permissions}. The paradigm of Pharos is mirrored in terrestrial traffic management, where vehicles from varying brands and drivers with varying experience can independently control their vehicles under the macro-coordination of traffic lights.

To realize this collaborative paradigm, we design and implement a system of the same name, \textbf{Pharos}. The system establishes global situational awareness by fusing lightweight UAV telemetry data with dynamic pedestrian states, where the latter can conceptually be acquired through crowdsourced IoT networks in real-world deployments. Driven by this, Pharos employs Multi-Agent Reinforcement Learning (MARL) for collaborative airspace coordination. During the training phase, Pharos encodes UAV observation features using the Artificial Potential Field (APF) to capture dynamic urban environments. To achieve global optimization with \textbf{safety priority}, we formulate a comprehensive reward mechanism that prioritizes \textbf{collision avoidance} and \textbf{human safety}, while incorporating \textit{flight progression as an auxiliary incentive}. Specifically, we innovatively introduce a \textbf{human fear factor} to quantify the psychological impact of low-altitude flights, incorporating physical distance, velocity, and direction vectors. Collision penalties strictly account for both inter-UAV collisions and stationary urban obstacles. During the inference phase, the trained model enables parallel execution across UAVs: at each time step, multiple UAVs perform \textit{inference in parallel} to determine their non-overlapping, \textbf{exclusive spaces}. Finally, we introduce \textbf{spatial entropy} in simulations to effectively evaluate their space utilization efficiency within the constrained airspace.

The main contributions of this paper are as follows: 

\textbf{(1)} We propose Pharos, a collaborative airspace management solution for multi-UAV coordination that bridges centralized control and distributed local perception, thereby reducing potential UAV collisions and human safety concerns. 

\textbf{(2)} We introduce a quantifiable human fear factor to model the psychological impact of low-altitude UAV operations, and integrate it into the reward mechanism of the MAPPO algorithm to perform parallel inference of the UAV exclusive spaces. 

\textbf{(3)} We implement the namesake Pharos simulation system using real-world urban data. Comprehensive experiments validate its accuracy, effectiveness, and performance in low-altitude space coordination.

\section{Related Work}
\subsection{Airspace coordination for UAVs}
For current research on UAV airspace coordination, we summarize three core methods: \textbf{(1)} \textit{The centralized planning and optimization method} mainly includes two categories: Multi-Agent Pathfinding (MAPF) and Numerical Optimization (NO). MAPF focuses on single-objective problems (such as the shortest path) and struggles with the complex weighting of multiple objectives and soft constraints. When objectives are weighted, state space and collision resolution become highly complex, reducing algorithm efficiency. NO employs mathematical functions to model constraints and flight objectives—such as optimal paths \cite{phung2021safety}, minimal energy consumption \cite{yao2025minimizing}, and economic load dispatch \cite{akbari2022cheetah} —transforming them into solvable optimization problems. To achieve these optimizations, Geometric-based methods are often used, employing coordinate transformations (e.g., geometric distances \cite{zhan2022geometric}, SE(3) \cite{seo2023geometric}, and graph topologies \cite{musil2022spheremap}) to describe UAV motion constraints. However, these centralized methods generally have high computational complexity, and their effectiveness at obstacle avoidance in dynamic environments is limited.

\textbf{(2)} \textit{The distributed reactive control method} is exemplified by the Artificial Potential Field (APF) method. These methods achieve collision avoidance by creating a virtual attractive field for target points and a virtual repulsive field for obstacles \cite{pan2021improved}. Compared to centralized methods, APF is faster computationally and better suited to dynamic environments. However, it is essentially a short-sighted optimization approach relying on instantaneous gradient descent, which can easily cause UAVs to get stuck in local minima (where attractive and repulsive forces balance), hindering their progress towards the destination. To alleviate this issue, researchers have proposed various improvements, such as integrating light transmission models by \citet{li2022large}, and introducing collinear force deflection angles by \citet{zhang2024research}.

\textbf{(3)} \textit{The collaborative optimization method} is represented by Multi-Agent Reinforcement Learning (MARL). Unlike APF's dependence on instantaneous gradients, MARL collaboratively optimizes strategies by learning gradients of long-term cumulative rewards, effectively handling global optimization problems and avoiding local minima. In recent years, MARL-based methods \cite{MAPPO} have made significant progress in collaborative UAV decision-making \cite{li2022collaborative}, effectively avoiding the local-minimum problem in APF by combining exploration and exploitation strategies.
 
%In recent years, MARL-based methods have made significant progress in UAV collaborative decision-making \cite{li2022collaborative,dai2022multi}. By combining exploration and exploitation mechanisms, they successfully overcome the inherent local minima issue of APF. Consequently, MARL, with its long-term decision-making and collaborative learning capabilities, becomes a powerful approach to achieving efficient and reliable airspace management.

\subsection{Human-Machine Coexistence Safety}

As the coexistence of humans and machines in various environments deepens, ensuring human safety while enabling intelligent machines to work efficiently becomes a critical issue. The concept of human safety in human-machine interactions was first introduced in Asimov's Three Laws of Robotics and the ISO 10218-1 Four Principles \cite{iso2012robots}. In fields such as industrial robotics \cite{pereira2022improving}, and autonomous driving \cite{liao2024bat}, the physical scope of human-machine interaction is typically limited to relatively small, enclosed, and mostly two-dimensional spaces. However, the impact of UAVs operating in three-dimensional spaces on human safety remains a significant challenge. Although \citet{van2023increasing} developed a Control Barrier Function (CBF) to enhance human-perceived safety in interactions between a single UAV and a single human, prioritizing human safety in complex, dynamic environments with multiple UAVs and humans remains to be fully addressed.

\section{System Model and Problem Formulation}
Sec. \ref{sec:space_model_sec} will introduce the system space model, including the urban setting model and the UAV exclusive space model. And Sec. \ref{sec:problemFormulation} will formalize the problem of the optimization objectives.

\subsection{Space Model}\label{sec:space_model_sec}
There are two distinct technical approaches for creating the space model to coordinate exclusive, safe, and available space for multiple UAVs in the urban environment. The first approach treats space as a continuous resource, allowing optimal coordination based on the flight needs of UAVs. This method minimizes coordinated space but entails high computational complexity for both space coordination and conflict detection. In contrast, the second approach divides space into numerous independent, fine-grained, discrete units, where each UAV receives an exclusive cuboid slice defined by integer multiples of these units. Although this method may lead to some space wastage, it significantly reduces computational demands compared to the continuous model. Consequently, Pharos adopts the discrete space model for UAV airspace management.

%Through relevant experiments (the experiment results are available in the submitted source code), we have found that the collision rate based on continuous space is 3 orders of magnitude higher than that based on discrete space due to the higher computational complexity. 
%, as demonstrated in Tab. \ref{tab:discrete_continuous_com} 
%\input{content/tab1}

\begin{figure}[t]
\centering
\includegraphics[width=0.5\textwidth]{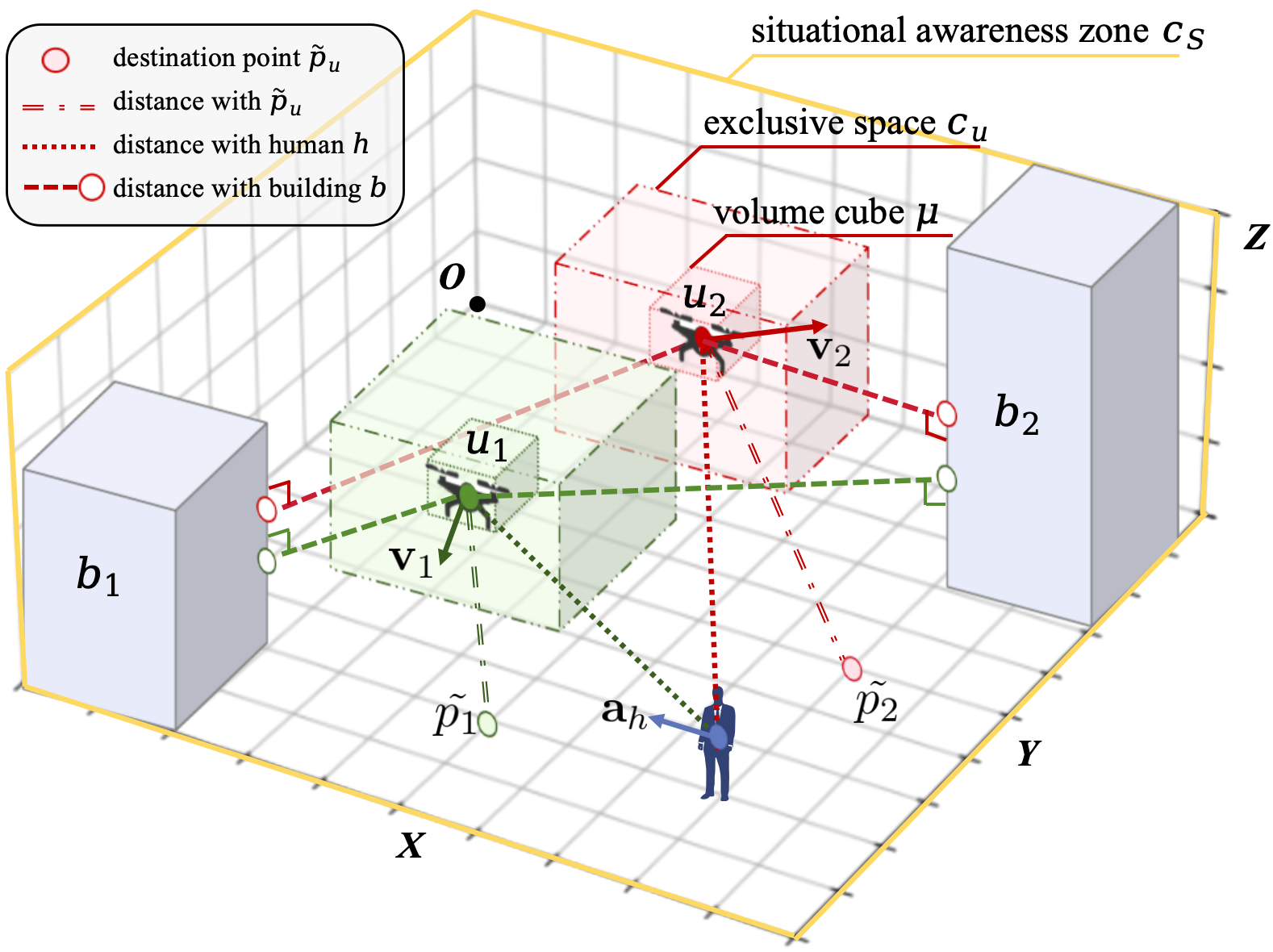} 
\caption{An illustration of the multi-UAV management in the complex urban space model.}
\label{fig:space_model}
\end{figure}

We use the space coordinate system \( S(x,y,z) \) shown in Fig. \ref{fig:space_model} to represent 3D airspace, which includes both moving humans and stationary buildings as obstacles. A set of humans \( H=\{h\} \) moves on the \( xOy \) plane, with each human \( h \) having a velocity \( \mathbf{a_h} \) and a gravity center at \( p_h(x,y,\overline{h}) \), where \( \overline{h} \) is a constant height. The stationary buildings are represented by \( B=\{b\} \). Following the design from Hu et al. \cite{hu2025research}, each building is modeled as a cuboid \( c_b (\{\varGamma_b^-,\varGamma_b^+ \}),\varGamma\in \{x,y,z\}\), based on its minimum and maximum projections on the \( x,y,z \) axes. This modeling method can be generalized to other types of stationary obstacles, such as sculptures or street lamps. 

There is a set of UAVs \( U=\{u\} \) operating in the airspace. To ensure safe operation, each UAV is coordinated an \textbf{exclusive space}, represented by a cuboid \( c_u(\{\varGamma_u^-,\varGamma_u^+ \}),\varGamma\in \{x,y,z\}\) (illustrated as the dashed box in Fig. \ref{fig:space_model}). This cuboid allows the UAV to customize its flight path without colliding with other UAVs or obstacles, ensuring that its exclusive space does not overlap with others. For UAV \( u \) with destination point \( \tilde{p_u} \) and current velocity \( \textbf{v}_u \), its current gravity center point is \( p_u(x,y,z) \). The UAV $u$ is regarded as a volumetric object rather than merely a particle point, and its volume is modeled by a cube of side length $\mu$ that must be entirely contained within the exclusive space. And the distance from UAV \( u \) to human \( h \) is represented by \( d(u,h) \), calculated as the Euclidean distance between their gravity centers. The distance from UAV \( u \) to building \( b \) is \( d(u,b) \), defined as the shortest Euclidean distance from the UAV's gravity center \( p_u \) to the surface of \( c_b \). The current space \( S \) is treated as a global situational awareness zone \( c_S \). The data center Pharos performs the above space modeling (as shown in Fig. \ref{fig:space_model}) and UAV airspace management using stationary building data and situational data reported by UAVs and other mobile intelligent terminals.

% We compute the exclusive space using the discretized space granularity. Compared to the continuous approach in Tab. \ref{tab:discrete_continuous_com}, collision detection between UAVs is reduced from a complex geometric intersection test of continuous cuboids to a straightforward check for overlapping sets of discrete voxels. This significantly reduces computational overhead, which is critical for real-time applications. It is not denied that continuous space saves more space, but with higher complexity. We will set the discrete granularity fine enough to approximate the continuous space.

%改标题. 协作与控制解耦的空间管理
\subsection{Problem Formulation}\label{sec:problemFormulation}
%The dynamic airspace management of UAVs is to find a strategy to ensure human-UAV safety and flight accomplishment. 

%We define the optimization objectives of the dynamic UAV airspace management problem. 

UAV collision avoidance and human safety are the foremost priorities. Additionally, to ensure a comprehensive airspace management solution, we also include flight progression in our optimization objectives. Therefore, the optimization objectives for Pharos's airspace management problem are divided into three aspects: the collision avoidance quantified by the \textbf{collision penalty} to avoid collisions among UAVs and between UAVs and obstacles; the \textbf{human fear penalty} to alleviate the fear UAVs may instill in humans; \textbf{flight progression} is to prompt UAVs to reach their destinations as soon as possible instead of hovering. Pharos is designed to find the optimal strategy for the multi-UAV exclusive space coordination.

%惩罚分成两部分，第一部分I表示u和其他无人机之间的交集点集数量。
\subsubsection{Collision Penalty}
The potential collision objects for UAV \( u \) include the other UAVs in \( U \), humans \( H \), and buildings \( B \), as Equ. \ref{equ:collision_penalty}. 

\begin{equation}
      P_u^s= \frac{1}{2}\sum_{u'\in   U \setminus \{u\}} I_{u,u'} + \sum_{k\in H\cup B} A_{u,k}   
\label{equ:collision_penalty}
\end{equation}
where,
\[
\begin{aligned}
    &I_{u,u'}=\prod_{\varGamma \in \{x,y,z\} } \max(\min(\varGamma_u^+,\varGamma_{u'}^+)-\max(\varGamma_u^-,\varGamma_{u'}^-),0) 
    \\
    &A_{u,k} = \left\{
\begin{aligned}
    1, &\quad d(u,e)\leq  \mu/2\\
    0, &\quad \text{otherwise}
\end{aligned}\right. 
\end{aligned} 
\]
The penalty is divided into two parts.  \( I_{u,u'} \)  represents the number of intersection points between the exclusive spaces of \( c_u \) and the other UAV \( c_{u'} \). The coefficient 1/2 in \( P_u^s \) eliminates double counting of intersection volumes for all UAV pairs in Equ. \ref{equ:total_reward}. Collisions between \( u \) and moving humans \( H \), as well as stationary buildings \( B \), are classified as collisions with obstacles. We define \( A_{u,k} \) to indicate a collision occurs when the distance from UAV \( u \)'s gravity center \( p_u \) to obstacle \( k\) is less than half the side length $\mu$ of its volume cube, aligning with intuitive expectations. Specific distance calculations between \( u \) and \( h \), and between \( u \) and \( b \), follow the system model definition in Sec. \ref{sec:space_model_sec}.

\begin{figure}[t]
\centering
\includegraphics[width=0.42\textwidth]{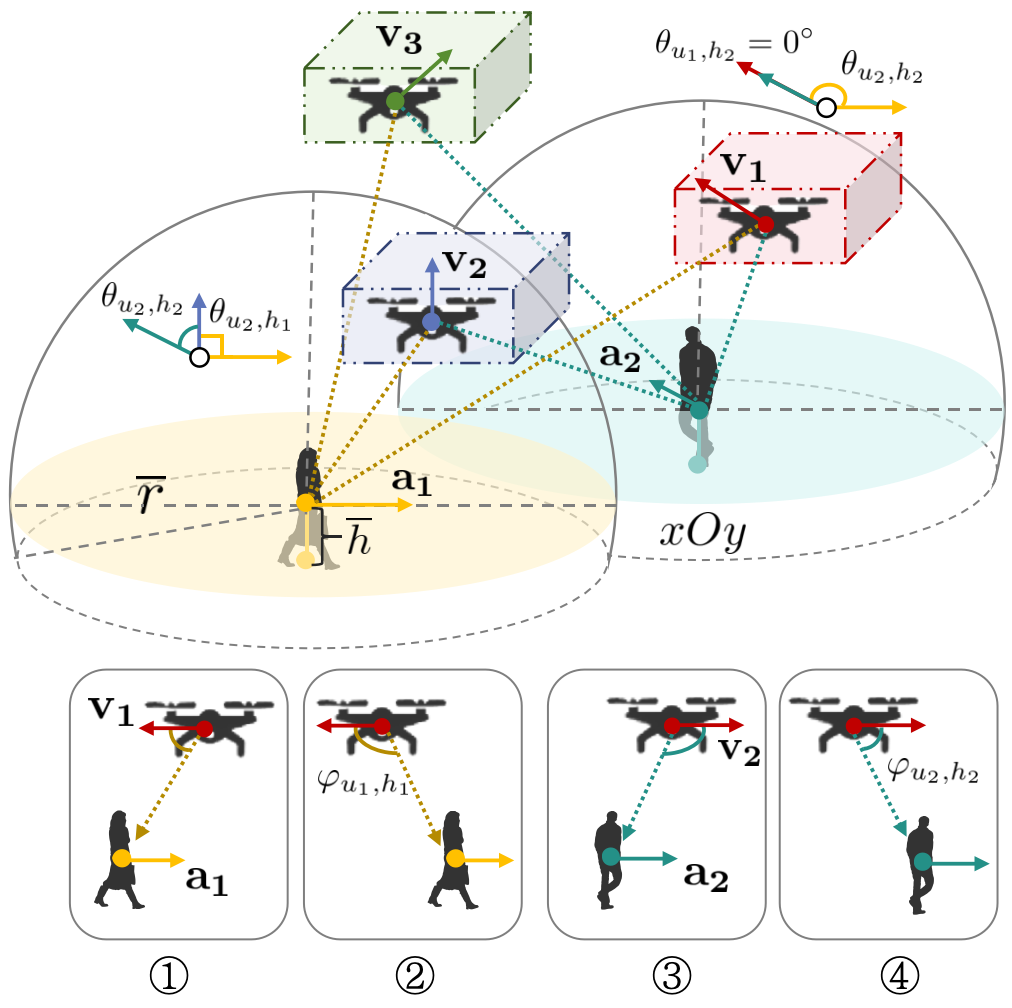} % Reduce the figure size so that it is slightly narrower than the column.
\caption{The human fear model with highlighting four specific cases \ding{172} to \ding{175} of UAVs' impact on humans, projected onto the \( xOz \) plane. Notably, Pharos operates with 3D vectors rather than 2D perspectives.}
\label{fig:fear_factor}
\end{figure}

\subsubsection{Human Fear Penalty}

We proposed the human fear factor to quantify the fear impact of UAV \( u \) on human \( h \), and the penalty value \(P_u^f\) is calculated as Equ. \ref{equ:human_fear_penalty}. The influence occurs only if the distance \( d(u,h) \) is within \(\bar{r}\) in Fig. \ref{fig:fear_factor}. And the angle \( \theta_{u,h} \) represents the angle between the UAV's velocity vector \(\mathbf{v_u}\) and the human's velocity vector \(\mathbf{a_h}\), while \( \varphi_{u,h} \) is the angle between \(\mathbf{v_u}\) and the vector from UAV pointing to the human. Fear increases when UAVs approach humans in case \ding{172} and decreases when they move away in case \ding{174}. Additionally, as the distance between the UAV and human increases in cases \ding{173} and \ding{175}, fear diminishes.

\begin{equation}
P_u^f = \sum_{h\in H}\left\{
\begin{aligned}
    &\frac{ (1-cos\theta_{u,h})\cdot \max(0,cos\varphi_{u,h})}{d(u,h)}, \bar{r}\geq d(u,h) \\
    & 0, \quad \text{otherwise}
\end{aligned}\right.
\label{equ:human_fear_penalty}
\end{equation}  

\subsubsection{Flight progression}
UAV $u$'s flight progression consists of two parts: the immediate bonus $R_u^l$ upon arriving at destination $\tilde{p_u}$ (see Equ. \ref{equ:flight_immediate_reward}), which is triggered when the distance from the gravity center $p_u$ to $\tilde{p_u}$ is below a threshold $\delta$. The second part is the progression bonus $R_u^g$ (see Equ. \ref{equ:flight_process_reward}), where $p'_u$ denotes the gravity center coordinate of $u$ at the previous timestamp. The flight progression is designed to encourage the UAV to fly forward and prevent it from hovering to avoid collisions.

\begin{subequations}
\begin{align}
&R_u^l = \left\{
\begin{aligned}
    1, &\quad d(p_u,\tilde{p_u})< \delta \\
    0, &\quad \text{otherwise}
\end{aligned}\right. \label{equ:flight_immediate_reward}\\
 &R_u^g=d(p'_u,\tilde{p_u})-d(p_u,\tilde{p_u}) \label{equ:flight_process_reward}
\end{align}
\end{subequations}

Thus, the final optimization objective for the UAV airspace management problem is represented as \(\mathcal{R}\) in Equ. \ref{equ:total_reward}. Here, \(\beta^s\), \(\beta^f\), \(\alpha^l\), and \(\alpha^g\) are the weight parameters for collision penalty, human fear penalty, immediate flight reward, and progression flight reward, respectively. Our goal is to maximize \(\mathcal{R}\) to coordinate exclusive spaces for UAVs for dynamic airspace management.

\begin{equation}
\max\mathcal{R}=\max\sum_{u\in U}(-\beta^s P_u^s-\beta^f P_u^f+\alpha^l R_u^l+\alpha^g R_u^g)
\label{equ:total_reward}
\end{equation}
%($\alpha_F$, $\alpha_D$, $\beta_C$,$\beta_H$) in $\mathcal{R}$

% The point $p$ within this coordinate system is defined as $p(x, y, z)$. Additionally, the UAV volume requires consideration, as it will increase computational complexity. the Euclidean distance between any two points $p$ and $q$ in space is $d(p,q)$. However, it has few applications in the existing research on collaborative management. A safety margin must be reserved for the UAV, and the overall safety of the fuselage space must be met during space management. We define the UAV volume as the \textbf{bounding box} (the first box outside the UAV in Fig. \ref{fig:space_model}), which is a cuboid $c$ and the minimum space that the UAV naturally needs to occupy in space $S$, such as when the UAV hovers. For example, taking the latest product of Dajiang Innovation, DJI Mavic 4 Pro (with propellers)\footnote{www.dji.com/cn/products/comparison-consumer-drones}, as an example, it has a length of $257.6 mm$, a width of $124.8mm$, and a height of $106.6 mm$.
%没有考虑无人机的体积 \cite{wang2024collaborative}

\section{Proposed Method} 
Given that this study mainly focuses on collaborative decision-making and optimization at the algorithm level, we assume that the Pharos system, based on a collaborative group framework with consistent goals and meeting the admission conditions, can reliably collect situation awareness data and distribute the exclusive space coordination results at each time step. The following Sec. \ref{sec:algoModel} formalizes the algorithm model as a Dec-POMDP, and Sec. \ref{sec:algo} introduces the MAPPO algorithm adopted.

%This section details the implementation method of the Pharos system, divided into: Sec. \ref{sec:algoModel} formalizes the algorithmic model as a Decentralized Partially Observable Markov Decision Process (Dec-POMDP), and Sec. \ref{sec:algo} introduces the Multi-Agent Proximal Policy Optimization (MAPPO) algorithm \cite{MAPPO} adopted to solve it. 

\subsection{Airspace Management Decision Model}\label{sec:algoModel}

The multi-UAV airspace management problem is essentially a collaborative Multi-Agent System (MAS), in which multiple UAVs must make collaborative decisions in a partially observable and dynamically changing environment to achieve global objectives. Therefore, we adopt the Decentralized Partially Observable Markov Decision Process (Dec-POMDP) as the formal algorithm framework of Pharos. This model accurately captures the characteristics of UAVs making decisions in parallel based on their observations, while the design of global states and shared reward functions aligns with the macro collaboration objectives of the Pharos center. This reflects the core concept of centralized collaboration and parallel execution from the perspective of the algorithmic framework.

We extracted the key components directly related to the Pharos airspace management problem from the standard Dec-POMDP model and represented them as a five-tuple  $<\mathcal{I},\mathcal{O},\mathcal{A},\mathcal{S},\mathcal{R}>$. A complete Dec-POMDP should also include standard components such as the state transition function, observation function, and discount factor. In this study, the state transition and observation functions are implicitly defined by the dynamics of the simulation environment. The discount factor $\gamma$ is a critical hyper-parameter, whose setting will be explained in Sec. \ref{sec:algo} and summarized in Tab. \ref{tab:simulation_setup}. We represent the entire airspace management cycle as $\mathcal{T} = \{t\}$. Pharos can handle the dynamic changes in the urban environment at each time step $t$ and coordinate appropriate exclusive spaces for UAVs, thereby ensuring reliable airspace management. The components of $<\mathcal{I},\mathcal{O},\mathcal{A},\mathcal{S},\mathcal{R}>$ corresponding to time step $t$ are elaborated in detail as follows:

\begin{figure}[t]
\centering
\includegraphics[width=0.5\textwidth]{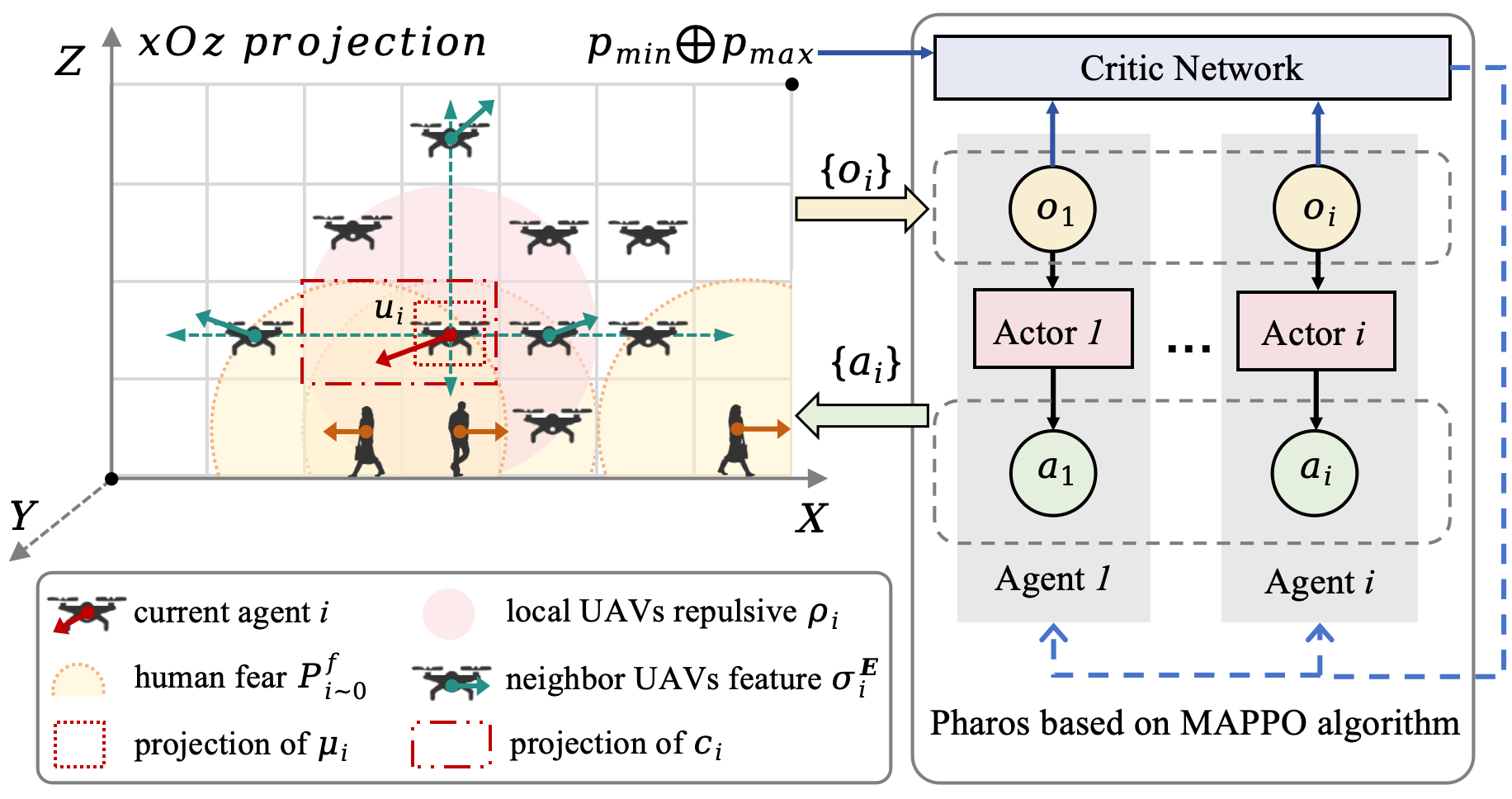} 
\caption{Pharos' core algorithm based on MAPPO.}
\label{fig:Pharos_algo}
\end{figure}

\textit{\textbf{Agent}}: $\mathcal{I}=\{i\}$ represents a definite set of UAV agents.

\textit{\textbf{Observation}}: $\mathcal{O}_t=\{o_i\}$ is the observation set of the agents at timestep $t$, with each $o_i$ formalized as: \[o_i = [id_i, N(p_i), N(\tilde{p_i}-p_i),m(p_i,\tilde{p_i}),\rho_i, \sigma_i^{\textbf{E}}, F_i]\]
where $o_i$ is a one-dimensional vector of size 24 and contains four types of features: 

(1) \textbf{Self feature}: $id_i$ is the integer identity number of agent $i$, and $N(p_i)=p_i/(p_{max}-p_{min})$ is the normalized vector of size 3 for the agent $i$'s gravity center position with respect to the coordinate system $S$, where $p_{max}$ and $p_{min}$ are the upper and lower bounds of $S$ respectively. $N(\tilde{p_i}-p_i)=(\tilde{p_i}-p_i)/(p_{\max}-p_{min})$ and $m(p_i,\tilde{p_i})$ are the normalized unit vector of size 3 and Manhattan distance value of $p_i$ and its destination $\tilde{p_i}$. 

(2) \textbf{partially UAVs repulsive feature}: We customized the \textit{rank} function to count the $\kappa$ UAVs closest to $i$ in $U$, which is used to calculate the partially UAV repulsive feature. Based on the definition of repulsive force in physics, the calculation between UAV $i$ and UAV $j$ is obtained, and the partially repulsive feature $\rho_i$ with respect to $i$ is expressed as follows.
\begin{equation}
    \rho_i=\sum_{j\in \text{rank}(U)_i^{\kappa}}(p_i-p_j)/d(i,j)^3
\label{equ:local_uav_density}
\end{equation}

(3) \textbf{Neighbor UAV velocity feature}: The neighbor velocity features of $i$ are $\sigma_i^{\textbf{E}}=\{\sigma_i^{\textbf{e}}\}$, with $\sigma_i^{\textbf{e}}$ formalized in Equ. \ref{equ:obs_neighbor_feature}. To explain the meaning of neighbor UAV, the six observation directions are defined as $\textbf{E} =\{\mathbf{e}\}=\{(0,0,e),(0,0,-e),(0,e,0),(0,-e,0),(e,0,0),(-e,0,0)\}$. Then the neighbor of $i$ in $\mathbf{e}$ is the nearest UAV $j$, that is $p_j= p_i+ \lambda \mathbf{e}$ (take $\lambda$ as the first positive integer that makes this expression hold). 
%\lambda=2在本文中
\begin{equation}
\begin{aligned}
&\sigma_i^{\textbf{e}} =\left\{
\begin{aligned}
&0.5+\varepsilon\cdot \text{sign}((\mathbf{v}_j - \mathbf{v}_i)\cdot \mathbf{e})/\lambda, & &p_j= p_i+ \lambda \mathbf{e},\\
&0, & &\text{otherwise,} 
\end{aligned}
\right. \\
&\text{where}, \ \lambda\in \mathbb{N}^+, \ \varepsilon\in[0,0.5].\\
\end{aligned}
\label{equ:obs_neighbor_feature}
\end{equation}
Given a neighbor \( j \) in direction \( \mathbf{e} \), \(\sigma_i^{\mathbf{e}}\) computes the relative velocity feature between agents \( j \) and \( i \) along \( \mathbf{e}\). This feature is inversely proportional to the distance coefficient \( \lambda \) between them. Geometrically, a positive relative velocity feature indicates that neighbor \( j \) is flying closer, thereby increasing \(\sigma_i^{\mathbf{e}}\), while a negative feature signifies that \( j \) is flying away, causing \(\sigma_i^{\mathbf{e}}\) to decrease.

(4) \textbf{Human fear prediction feature}: With the six observation directions mentioned above, the operation of the UAV can also hover at the current position. Therefore, the prediction of human fear features includes seven values, which are concatenated using Equ. \ref{equ:human_fear_penalty} as follows. 
\begin{equation}
    F_i=P_i^f \oplus P_{i \sim \mathbf{1}}^f\oplus... \oplus P_{i \sim \mathbf{e}}^f
\label{equ:obs_fear}
\end{equation}
where $P_i^f$ represents the fear that occurs to humans when $i$ hovers, and $P_{i\sim \mathbf{e}^f}$ represents the fear that occurs to humans if $i$ moves one step in the direction of $\mathbf{e}$.

\textit{\textbf{Action}} $\mathcal{A}_t=\{a_i\}$ indicates that the agent set has a definite set of actions at the timestep $t$. The geometric meaning of the action set is the direction and range selection for exclusive spaces. Therefore, in the discrete 3D space, six directions $\mathbf{E}$ (the same as the neighbor UAVs) need to be considered, as well as an additional action of the UAV hovering, that is, $\mathbf{E}=\mathbf{E}\cup\{(0,0,0)\}$. Then $a_i$ is denoted as follows:
\[a_i=[a_i^0,...,a_i^\textbf{e}],\sum_{a_i^\mathbf{e}\in a_i}a_i^\mathbf{e}=1,a_i^\textbf{e}\in \{0,1\}.\]
Each agent is coordinated the exclusive space $c_i$ in the direction $\mathbf{e}$ of $a_i^\mathbf{e}=1$ according the observation information $a_i$, and the calculation  $c_i$ is formalized as follows:
\begin{equation}
c_i=(\{\varGamma_i^--e/2,\varGamma_i^++e/2\})
\label{equ:exclusive_space_cal}
\end{equation} 
where,
\[
\begin{aligned}
\varGamma_i^-=&\min(\varGamma_i,\varGamma'_i),\ \varGamma_i^+=\max(\varGamma_i,\varGamma'_i),\varGamma \in \{x,y,z\},\\
p'_i&(x'_i,y'_i,z'_i)=p_i(x_i,y_i,z_i)+\sum_{a_i^\mathbf{e}\in a_i}a_i^\textbf{e}\cdot\mathbf{e}.
\end{aligned}\]
In Equ.\ref{equ:exclusive_space_cal}, $p'_i$ is the next position coordinate of UAV $i$ calculated by the action  $a_i$ and the current position $p_i$. The exclusive space $c_i$ (as Fig.\ref{fig:Pharos_algo}) is obtained by forming a cuboid with coordinates $p_i$ and $p'_i$ and extending the boundary outward by a length of $e/2$. The UAV's operation within $c_i$ can avoid potential collisions and eliminate human fear.

\textit{\textbf{State}} The environment state at timestep $t$ is a one-dimensional vector (with the size of $6+24*|\mathcal{I}|$) as:
\[\mathcal{S}_t=[p_{min},p_{max},o_1,...,o_i].\]
Where $p_{min}$ and $p_{max}$ are the coordinate system information and $\{o_i\}$ are the agents' observation information.

%This state will also serve as the input of the critic network for evaluating the actor networks' policy.
\textbf{\textit{Reward}} The reward function provides feedback on the quality of agents' actions, facilitating a balance between exploring new behaviors and exploiting known effective ones. We define the reward function \(\mathcal{R}_t\) in Equ. \ref{equ:total_reward} shared by all agents to guide UAVs in optimizing the coordination of exclusive spaces. 

%in the context of cooperative game theory,
Within the framework of Dec-POMDP, the multi-UAV airspace management problem is considered an \textit{NEXP-complete problem} \cite{bernstein2002complexity}, posing theoretical challenges. However, due to the \textit{finite-horizon} \cite{oliehoek2016concise} of the current problem, there exists at least one optimal joint policy that satisfies the following conditions: The finite time step set $\mathcal{T}$, observation set $\mathcal{O}_t$, action set $\mathcal{A}_t$ and state set $\mathcal{S}_t$ of UAVs; The bounded reward function $\mathcal{R}_t$ consist of collision penalty, human fear penalty, and flight reward; the system’s state transitions adhere to the Markov property. Thus, our goal is not to find a precise optimal solution but rather to have multiple UAVs collaborate to approximate the global optimal, safe, and efficient airspace coordination strategy.

\begin{algorithm}[tb]
\caption{Pharos: dynamic airspace management algorithm for multiple UAVs based on MAPPO} 
\label{alg:algorithm}
%\raggedright
%\textbf{Input}: UAVs $U$, humans $H$, buildings $B$, 3D-space $S$\\
%\textbf{Output}: result set of multi-UAV exclusive spaces $\{c_{i}|i\in U\}$
\begin{algorithmic}[1] %[1] enables line numbers
\REQUIRE UAVs $U$, humans $H$, buildings $B$, 3D-space $S$
\ENSURE result set of multi-UAV exclusive spaces $\{c_{i}|i\in U\}$
\STATE init the critic $V_{\omega}$, actors $\{\pi_{\vartheta_i}|i\in \mathcal{I}\}$
\WHILE{$ep\leq episode$}
\STATE init environment $t=0,\mathcal{S}=\mathcal{S}_0,\mathcal{D}=[\ ]$
\WHILE{$t<=\mathcal{T}$}
\STATE get observation set from env $\mathcal{O}_t=\{o_i\}$
\STATE execute each actor: $\mathcal{A}_t\{a_i\}\leftarrow \mathcal{O}_t\{o_i\}$ 

\STATE execute Equ. \ref{equ:exclusive_space_cal}: $\{c_i\} \leftarrow \mathcal{A}_t\{a_i\}$
\STATE take $\mathcal{A}_t$ to step env: $\mathcal{S}_{t+1},\mathcal{R}_t \leftarrow \mathcal{S}_t$

 \STATE store $(\mathcal{S}_t, \mathcal{A}_t, \mathcal{R}_t, \mathcal{S}_{t+1})$ in replay buffer $\mathcal{D}$

%\STATE sample a ramdon batch $srd=\{\hat{\mathcal{S}_t}, \hat{\mathcal{A}_t}, \hat{\mathcal{R}_t}, \hat{\mathcal{S}}_{t+1}\}$
\ENDWHILE
\STATE gradient update the critic network $V_{\omega}$ by loss function $loss_{\omega}=\frac{1}{\mathcal{T}}\sum_{t \in \mathcal{T}}(G_t-V_{\omega}(\mathcal{S}_t))^2$   

\STATE gradient update each actor $\pi_{\vartheta_i}$ by $loss_{\vartheta_i}=\frac{1}{\mathcal{T}}\sum_{t \in \mathcal{T}} $\\$(G_t-V_{\omega}(\mathcal{S}_t))\cdot\min[ prob(a_i),\text{clip}(\text{prob}(a_i),1-\varsigma,1+\varsigma)]$, $\text{prob}(a_i)=\pi_{\vartheta_i}(a_i)/\pi^{ep-1}_{\vartheta_i}(a_i)$ 
\ENDWHILE
\end{algorithmic}
\end{algorithm}
\subsection{Airspace Management Algorithm for UAVs}\label{sec:algo}
In the Dec-POMDP modeling for multi-UAV airspace management, we employ the Multi-Agent Proximal Policy Optimization (MAPPO) algorithm \cite{MAPPO} for obtaining approximate solutions. MAPPO adopts a centralized training and decentralized execution paradigm. During the training phase, global information is used to optimize the joint policy of multiple UAVs. In the execution phase, each UAV independently makes decisions based on its partial observations and subsequently plans detailed paths. This paradigm aligns perfectly with the architecture of the Pharos system, which strikes a balance between centralized control and distributed local perception paradigms, ensuring efficient collaboration and scalability at the algorithmic level. Moreover, to validate the reasonableness of our algorithm selection, we conducted comparative experiments with typical multi-agent reinforcement learning algorithms such as HAPPO and HATRPO. The relevant results and analysis are detailed in Sec. \ref{sec:simulation_result}.

%Screenshot of Pharos visualization system and two test cases of collision avoidance capability. And the complete dynamic visual effect is demonstrated in the video of the anonymized code, available at
\begin{figure*}[t]
\centering
\includegraphics[width=0.98\textwidth]{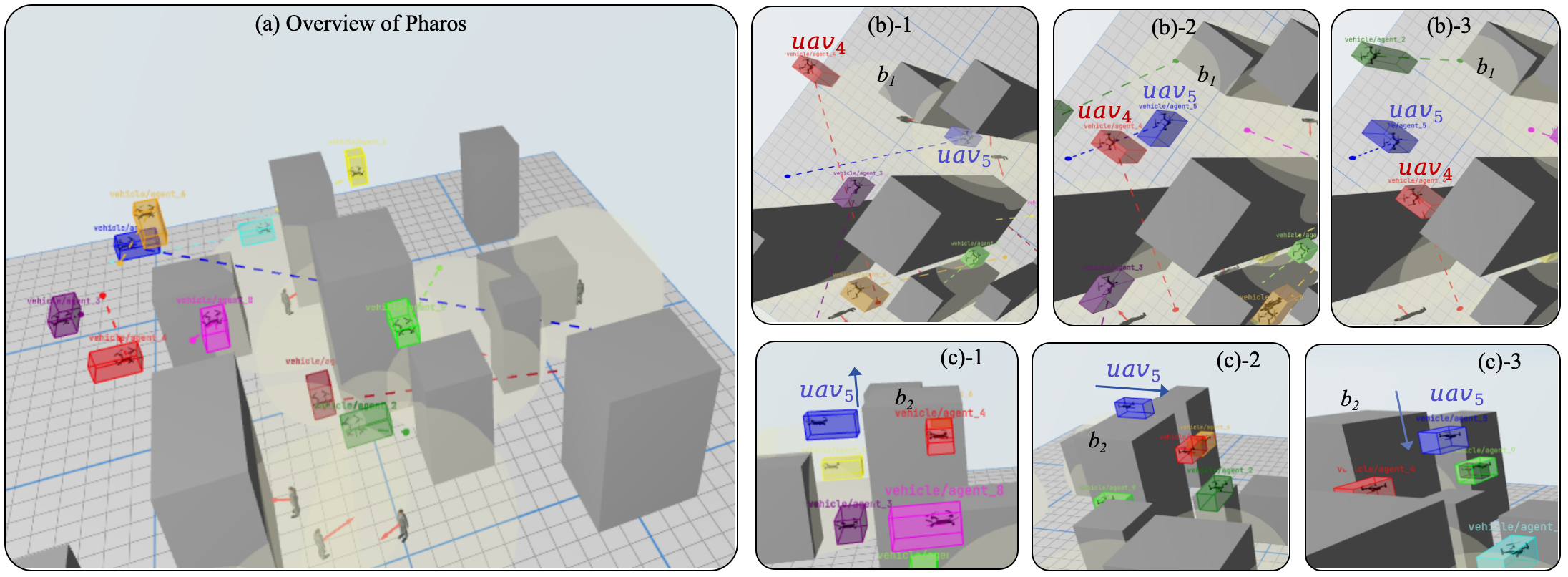} 
\caption{(a) Overview of Pharos visualization system. (b)(c) Snapshots of two test cases for system capabilities. Full dynamic effect is shown in the anonymized video at \textit{\url{https://github.com/pharos-anonymized/source-code/blob/master/visualization.md}}.}
\label{fig:collision_impact}
\end{figure*}

As illustrated in Fig. \ref{fig:Pharos_algo}, the Pharos system built on the MAPPO algorithm comprises two core components: a set of actor networks $\{\pi_{\vartheta_i}\}$, each corresponding to one UAV, and a global shared critic network $V_{\omega}$ evaluating the performance of actors' policies. Both the actor and critic networks are implemented as multi-layer perceptrons (MLPs). Each actor takes the partial observation $o_i$ of its associated UAV as input and outputs a probability distribution over possible directions for selecting the exclusive space. To conform to the discrete action set $\mathcal{A}_t$ defined in the decision model, the action $a_i$ with the highest probability is assigned a value of 1, while all others are set to 0. The critic receives the global state $\mathcal{S}_t$ as input and produces a value estimate $V_\omega$, which evaluates the quality of the current joint policy and guides the gradient-based update of all actor networks.

The Pharos algorithm, based on MAPPO, optimizes the policy $\pi_{\vartheta_i}$ of each UAV actor network through the policy gradient method. This optimization aims to maximize the long-term cumulative return $G_t$, where $G_t$ is defined as the sum of the reward $\mathcal{R}_t$ at the current timestamp and the discounted cumulative return at the next timestamp $G_{t+1}$. The algorithm updates the policy parameters by calculating the expected policy gradient $\nabla_{\vartheta_i} J(\vartheta_i)$, which represents the direction in which the long-term return $G_t$ changes with the policy parameters $\vartheta_i$. The formal calculation of our adopted algorithm MAPPO \cite{MAPPO} is as follows:
\[
\nabla_{\vartheta_i} J(\vartheta_i) = \mathbb{E}_{ \pi_{\vartheta_i}} \left[ \nabla_{\vartheta_i} G_t \right],G_t=\mathcal{R}_t+\gamma \cdot G_{t+1}.
\]
As the gradient gradually converges, the Pharos system eventually approaches a team-optimal solution in the cooperative multi-UAV game. That is, the policies of all UAV actors reach a local optimal state, and any unilateral change of policy by a single UAV cannot increase the global reward $\mathcal{R}_t$. The complete algorithmic procedure of Pharos is summarized in Algorithm \ref{alg:algorithm}, which can coordinate exclusive spaces for multiple UAVs in complex urban environments with stationary buildings and moving humans, while ensuring the safety of both humans and UAVs.

\section{Simulation}
We have developed a multi-UAV simulation visualization system for Pharos in a complex urban environment, as an overview in Fig. \ref{fig:collision_impact}(a) and anonymized source code available at \textbf{\textit{\url{https://github.com/pharos-anonymized/source-code/tree/master/pharos-visual-3d}}}. Detailed experiments and analysis are as follows.

\subsection{Simulation Setup}\label{sec:setup}
For the simulation dataset, we obtained stationary building data for a 155m × 135m rectangular area in Shanghai, China, representing a typical complex intersection, from OpenMap (OSM) \cite{OSM} and modeled the buildings based on designs by \citet{hu2025research} to construct the Pharos spatial map. Pharos also stochastically generates moving humans with diverse velocities and trajectories to simulate real-world dynamic interactions. During inference, the static layout remains fixed while all dynamic configurations, including UAV destination positions and human movements, are randomized. For UAV flight simulation, we have not preset any specific planned flight routes. We only require that the UAVs be able to reach the ``next exclusive space" at the next time step. This minimal constraint design allows users to retain complete freedom in route customization for actual UAV operation. 

The relevant hyperparameters for the implementation are detailed in Tab. \ref{tab:simulation_setup}. Among them, based on safety priority, the reward weights are: the paramount penalty for collision avoidance, alongside a cumulative penalty for human psychological impact, and the combined incentive for flight progression (including immediate bonus and progression bonus). This ensures that flight progression serves merely as an auxiliary objective, subject to safety constraints. According to the statistical data, various types of UAVs have a maximum level flight speed of 100 km/h \cite{state2023}, and a minimum end-to-end control latency of 50 ms \cite{caac2018}, indicating that a UAV can move up to 1.4 m per communication time step. Consequently, we establish the granularity of the discretized spatial model at the meter scale. All simulation experiments were run on a server equipped with an AMD EPYC 9554 CPU (128 cores), 251GB of RAM, and an NVIDIA RTX A6000 GPU (48GB VRAM).

\begin{table}[htb]
\centering
	\caption{Hyperparameter settings in the simulation.}
	\label{tab:simulation_setup}
	\begin{tabular}{cc}\toprule
		\textbf{Hyperparameter}& \textbf{Value}
        \\ \midrule
	space model: $\delta,\mu,e$ &0.02, 0.5, 1\\
human fear model: $\bar{h},\bar{r}$&1.7, 5.0\\
observation features: $\kappa,\varepsilon,\lambda$ & $|U|-1$, 0.3, 2\\   
reward weights: $\beta^s,\beta^f,\alpha^l,\alpha^g$ & 100.0, 2.5, 3.0, 1.0 \\
loss function: $\gamma ,\varsigma $ & 0.95, 0.2 \\
learning rate: actor, critic& 0.005, 0.002\\
episode step, batch size & 5e6, 100\\ \bottomrule
	\end{tabular}
\end{table}

\begin{figure*}[htb]
\centering
\includegraphics[width=0.9\textwidth]{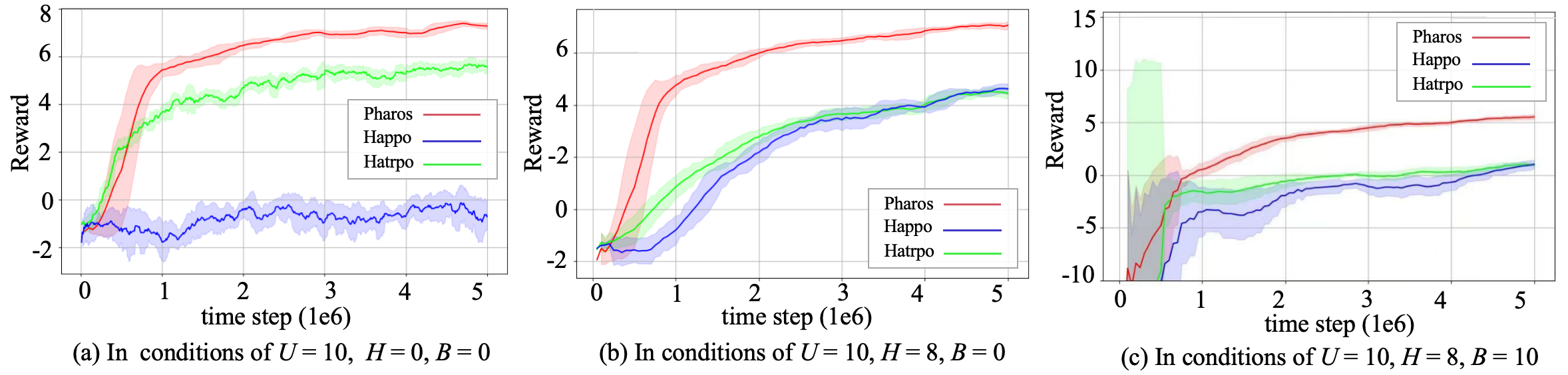} 
\caption{Comparison of training convergence with other typical MARL algorithms under different numbers of UAVs (\textit{U}), humans (\textit{H}), and buildings (\textit{B}).}
\label{fig:compare_diff_rl}
\end{figure*}

\subsection{Experimental Baselines}\label{sec:baseline}
The description of the experimental baseline algorithms and the reasons for the selections are as follows. 

\textbf{Ipopt}: The solver Ipopt (Interior Point Optimizer) is proposed by Stefan \cite{ipopt,coin-or-mumps}, which is an open-source mathematical optimization toolkit based on the interior point method and widely used for solving constrained optimization problems. It provides a continuous solution for Pharos' global optimization objective $\mathcal{R}$, serving as the accuracy benchmark for Pharos' discrete model.

\textbf{HAPPO and HATRPO}: HAPPO (Heterogeneous-Agent Proximal Policy Optimisation) and HATRPO ( Heterogeneous-Agent Proximal Policy Optimisation) are both multi-agent policy optimization algorithms proposed by \citet{kuba2021trust} in the same paper. We employed the same hyperparameter configuration as the MAPPO algorithm (see Tab. \ref{tab:simulation_setup}) to compare the performance of different MARL algorithms in the multi-UAV airspace management problem.

\textbf{A-star}: A-star is a classic benchmark in the field of path planning, and \citet{farid2022modified} have applied it to the path planning of UAVs. Although Pharos primarily addresses the issue of multi-UAV airspace collaborative management, it can also guide flights by coordinating exclusive spaces for UAVs. We will take the optimal path generated by A-star as the comparison benchmark and, under the same safety constraints, verify that Pharos can achieve planning efficiency similar to traditional methods in complex environments.

\subsection{Simulation Results and Analysis}\label{sec:simulation_result}
In this subsection, we conducted detailed experiments to verify and analyze the accuracy, effectiveness, and performance of the system.

\begin{figure}[htb]
\centering
\includegraphics[width=0.48\textwidth]{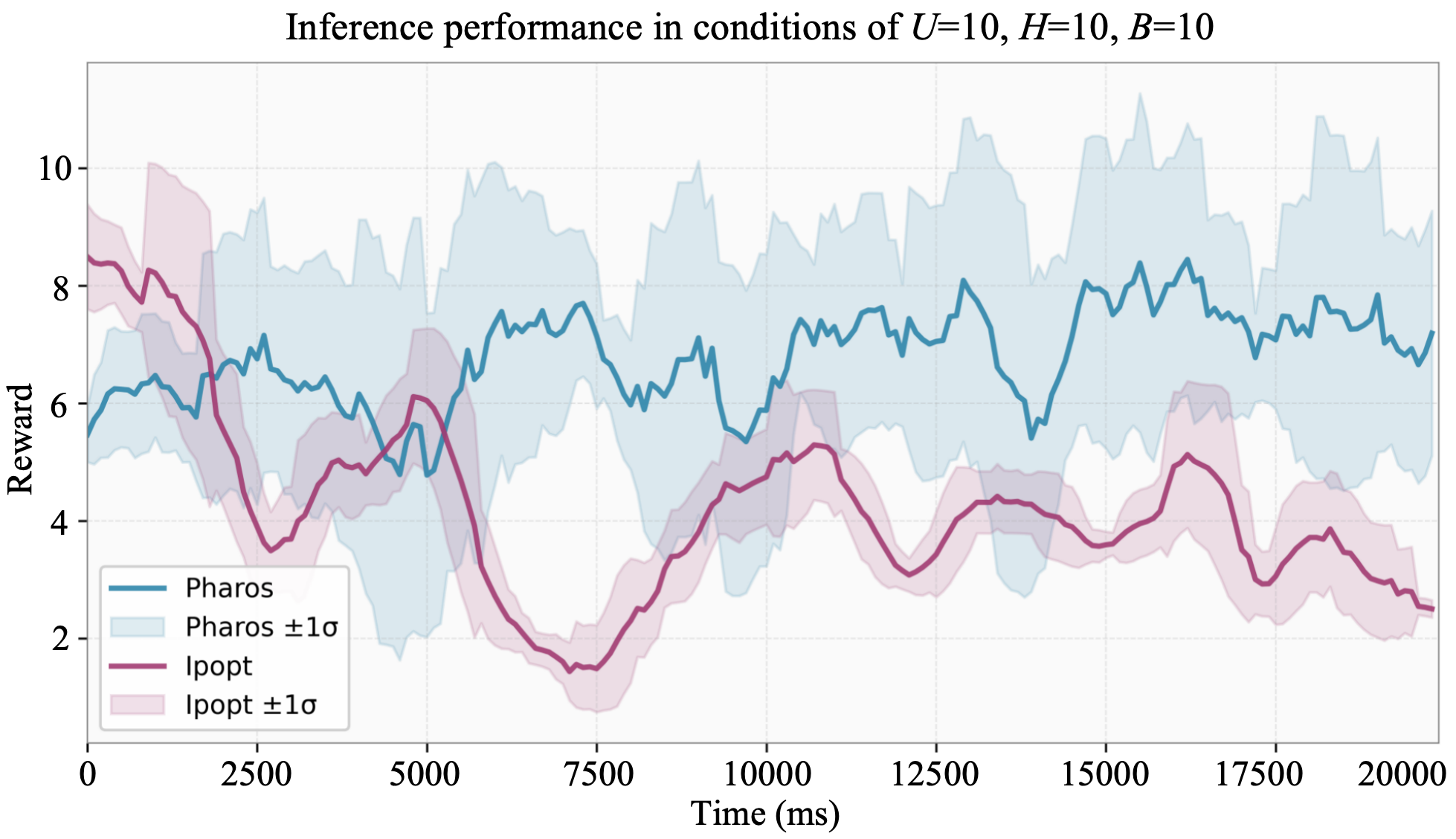} 
\caption{Comparison of inference performance between Pharos and Ipopt solver over 200 steps (100 ms/step).}
\label{fig:inference_test}
\end{figure}

\subsubsection{\textbf{Global optimization performance}}\label{sec:simualtion_global_opt}

%ipopt: 42.2950, Pharos: 67.9235 
%求解器没有长程规划，所以后半段奖励下降
%到最后一个时刻，RL最低的也比求解器高

To verify the solution accuracy and convergence performance of Pharos for the global optimization objective $\mathcal{R}$, we compared it with two types of baseline methods: one based on the mathematical optimization solver Ipopt, and the other using advanced MARL algorithms HAPPO and HATRPO. 

Firstly, we compared Pharos and Ipopt for the same optimization objective $\mathcal{R}$. During the inference phase, Pharos used a pre-trained model in an environment with 10 buildings, 10 humans, and 10 UAVs. As shown in Fig. \ref{fig:inference_test}, the reward value curves over time for both methods indicate that, despite some fluctuations during the initial inference phase, Pharos remains stable and significantly outperforms Ipopt most of the time, especially in the latter half of the inference, with an average reward value 60.59\% higher. This occurs because while Ipopt achieves exact mathematical optimality for a single time step, it lacks long-term planning capabilities.

Additionally, we compared Pharos with the HAPPO and HATRPO algorithms to test the model convergence during the training phase. All three used the same reward function and hyperparameters listed in Tab. \ref{tab:simulation_setup} and were trained in three scenarios: UAV-only, UAV-human, and UAV-human-building. As shown in Fig. \ref{fig:compare_diff_rl}, Pharos exhibits fast and stable convergence across all scenarios (a)-(c). Moreover, Pharos achieved the highest reward value after the fastest convergence in all scenarios, demonstrating that Pharos, based on the MAPPO algorithm, can better find a set of approximately optimal UAV collaborative management policies under our problem constraints.

\subsubsection{\textbf{Collision avoidance and flight progression}} 
%Following the design paradigm of centralized training and decentralized execution, during the training phase, the Pharos center conducts collaborative strategy learning for UAV agents based on global information. However, in actual deployment, each UAV relies on local observations for independent decision-making. 

To evaluate the Pharos's abilities of collision avoidance and flight progression, we designed two test cases and conducted visual analyses based on a pre-trained model comprising 10 buildings, 8 humans, and 10 UAVs, as shown in Fig. \ref{fig:collision_impact}(b)(c). 

\textit{Airspace coordination of UAVs at Intersection}. This test case replicates the scene from Fig. \ref{fig:intro_overview}to verify the system's ability for airspace coordination in shared airspace. The experimental results, presented in timeline (b)-1 to (b)-3 in Fig. \ref{fig:collision_impact}(b), outline three stages of their encounter: before the encounter, during safe collision avoidance, and as they keep approaching their destinations. Although building \( b_1 \) acts as a direct obstacle between $uav_4$ and $uav_5$, the UAV agents can obtain the appropriate exclusive spaces based on the pre-trained policies, operate safely within them, and avoid collisions. Furthermore, the visual trajectories confirm that the UAVs do not fall into deadlocks; driven by the progression incentives, they immediately resume cumulative progression toward their respective destinations post-encounter.

\textit{Obstacle Adaptive Detour}. This test focuses on the autonomous adaptability of UAVs when flying through tall buildings. As shown on the timeline (c)-1 to (c-2) in Fig. \ref{fig:collision_impact}(c), \(uav_5\) can leverage space coordination policy learned during model training to adjust its route in real-time during actual decision-making, achieving smooth and safe obstacle avoidance. Importantly, the flight progression reward can guide the UAV to maintain its forward momentum after bypassing the building, preventing any hovering or stagnation. This result shows the system’s environmental adaptability and sustained execution in complex obstacle environments.

\subsubsection{\textbf{Human fear impact}}
To evaluate the psychological impact on humans caused by UAV operations under different algorithms, we use Equ. \ref{equ:human_fear_penalty} to calculate the fear value experienced by humans due to nearby UAVs. We use the average fear values for UAV-human pairs as a comparative indicator of psychological fear impact among the algorithms. The constant value of $d(u,h)$ in Equ. \ref{equ:human_fear_penalty} is defined as the unit distance of 1 in the discrete spatial model to ensure absolute safety between UAVs and humans, meaning no collisions occur. Thus, the range of average fear values for UAV-human pairs is [0,2], where a smaller average fear value indicates that the algorithm is more human-friendly. For the Pharos algorithm, we utilize pre-trained models with a centralized approach for various numbers of UAVs (10 buildings, 8 humans). During the parallel execution of UAVs, we record the inference results at each time step. In contrast, with the Ipopt solver and the A-star algorithm, we synchronously collect the human fear values corresponding to each time step through step-by-step solving and planning processes in the same simulation environment.

As shown in Tab. \ref{tab:human_fear}, Pharos reduces human fear by an average of 52.72\% compared to A-star. However, we observed that compared to Pharos, Ipopt shows even lower human fear. This is because of its lack of long-term planning (\textbf{as discussed in Sec. \ref{sec:simualtion_global_opt}}). In experiments, when the UAV encountered moving humans or high-rise buildings, Ipopt determined that the optimal reward function was for the UAV to remain hovering, rather than taking a detour towards its destination. This represents a suboptimal strategy. While this may appear to yield lower human fear, it often leads to numerous UAVs becoming stuck and failing to fly forward. This phenomenon is further illustrated in the overall reward trends in Fig. \ref{fig:inference_test}.

\begin{table}[t]
\centering
\caption{Results of human fear and space utilization}%over 200 inference steps
\label{tab:human_fear}
\renewcommand{\arraystretch}{1.1}
\begin{tabular}{ccccc}
\toprule
\multirow{2}{*}{\textbf{Method}}&\multirow{2}{*}{\textbf{Metric} }&\multicolumn{3}{c}{\textbf{UAV count}}\\

\cline{3-5}
& & \vphantom{10}10 & 20 & 30 \\
           \hline
\multirow{2}{*}{Ipopt}& h.f. &0.1720 &0.3953&0.3131\\
          & s.e. & 29.5083 & 54.3881 & 81.9306\\
           \hline
\multirow{2}{*}{A-star}& h.f.&1.0436  &1.9870&2.6175\\

 & s.e. & 48.9936 &93.8634 & 134.0800\\
\hline
\multirow{2}{*}{\textbf{Pharos}} &h.f. &0.4228 &0.9273 &1.4304\\
 & s.e. & 49.7410 & 95.9659 & 137.1851 \\
 \bottomrule
\end{tabular}
\begin{tablenotes}
\item \textbf{Note}: Average human fear (h.f.) and spatial entropy (s.e.) in different methods over 200 inference steps (100 ms/step).
\end{tablenotes}
\end{table}

\subsubsection{\textbf{Space utilization}} 
%是不是需要加个归一化处理
To verify the space utilization of the UAV collaborative management system Pharos, we introduce the concept of \textit{spatial entropy} in this subsection. Spatial entropy is a dedicated measurement metric used for post-hoc evaluation of airspace resource utilization, rather than being part of the main Pharos algorithm. Inspired by the widely applied information entropy \cite{liang2024information,liu2022analysis,he2024oscillatory}, which denotes a system's degree of disorder, we have extended this concept to spatial entropy to quantify the uniformity of UAV distribution within the airspace. Specifically, a lower spatial entropy value indicates a more concentrated distribution of UAVs (e.g., all clustered on hotspot flight routes), which usually implies uneven utilization of airspace resources and a high risk of congestion. Conversely, a higher spatial entropy value suggests a more widespread and uniform distribution of UAVs across the available airspace, symbolizing more efficient and thorough utilization of airspace resources..

To calculate the value of spatial entropy, we analogize the event of a single UAV occupying a specific spatial unit in the Pharos coordinate system to a random event in information theory, according to the original definition of information entropy (brief expression, $\mathcal{H}=-\sum \mathcal{P} \text{log}\mathcal{P}$) proposed by C. E. Shannon \cite{shannon1948mathematical}. By statistically analyzing the frequency of occupation of each spatial unit by all UAVs during the task execution time, we can obtain the probability distribution of occupation for each spatial unit. Ultimately, this leads to the spatial entropy representing the overall characteristics of UAV distribution within he airspace, the calculation formula shown as follows:

\begin{equation}
   \mathcal{H}=-\sum_{q\in S}\mathcal{P}_{q}\log(\mathcal{P}_{q}),\ \mathcal{P}_{q}=\sum_{t \in \mathcal{T}}\sum_{u \in U} \mathbb{I} (p_u^t=q)/|\mathcal{T}|
\label{equ:entropy}
\end{equation} 
where \( q \) is a coordinate \( q(x, y, z) \) in space \( S \), \( p_u^t \) denotes the coordinate of UAV \( u \) at time \( t \), and \( \mathbb{I} \) is an indicator function that is 1 when \( p_u^t = q \) and 0 otherwise. 

We continue to compare the pre-trained Pharos with the Ipopt and A-star algorithms. The experimental results of spatial entropy for the three methods under different UAV counts are presented in Tab. \ref{tab:human_fear}. The spatial entropy for Pharos was the highest, with increases of 70.82\% and 2.03\% compared to Ipopt and A-star, respectively. The spatial entropy for the two algorithms differs significantly. For Ipopt, as discussed in the previous section, the algorithm's tendency to hover when facing obstacles results in low human fear and consequently low entropy, as it does not occupy additional space. Conversely, the UAVs using A-star cumulatively move toward their destinations, disregarding the fear it may instill in humans, thereby resulting in relatively high spatial entropy.

%解释：求解器表现出来是原地不动，所以对人的恐惧影响小，但是熵低，不会再走别的格子；a star一定会动，即便有人和建筑物，但有可能会造成更多的人类恐惧。

% Ipopt：68.57+76.45++67.44
%A-star：1.53+2.24+2.32=

% 只考虑无人机的时候，不考虑人和建筑物以及无人机的大小，测试在不同无人机数量下，连续和离散空间性能的变化，收敛情况对比 (实际操作看下来，连续空间在不考虑人和建筑物的情况下，收敛速度快，但是考虑了人和无人机又无法收敛，这个展示pass吧)

\subsubsection{\textbf{Complexity analysis}}
%具体分析一下复杂度多高：分成训练和推理

\begin{figure}[htb]
\centering
\includegraphics[width=0.45\textwidth]{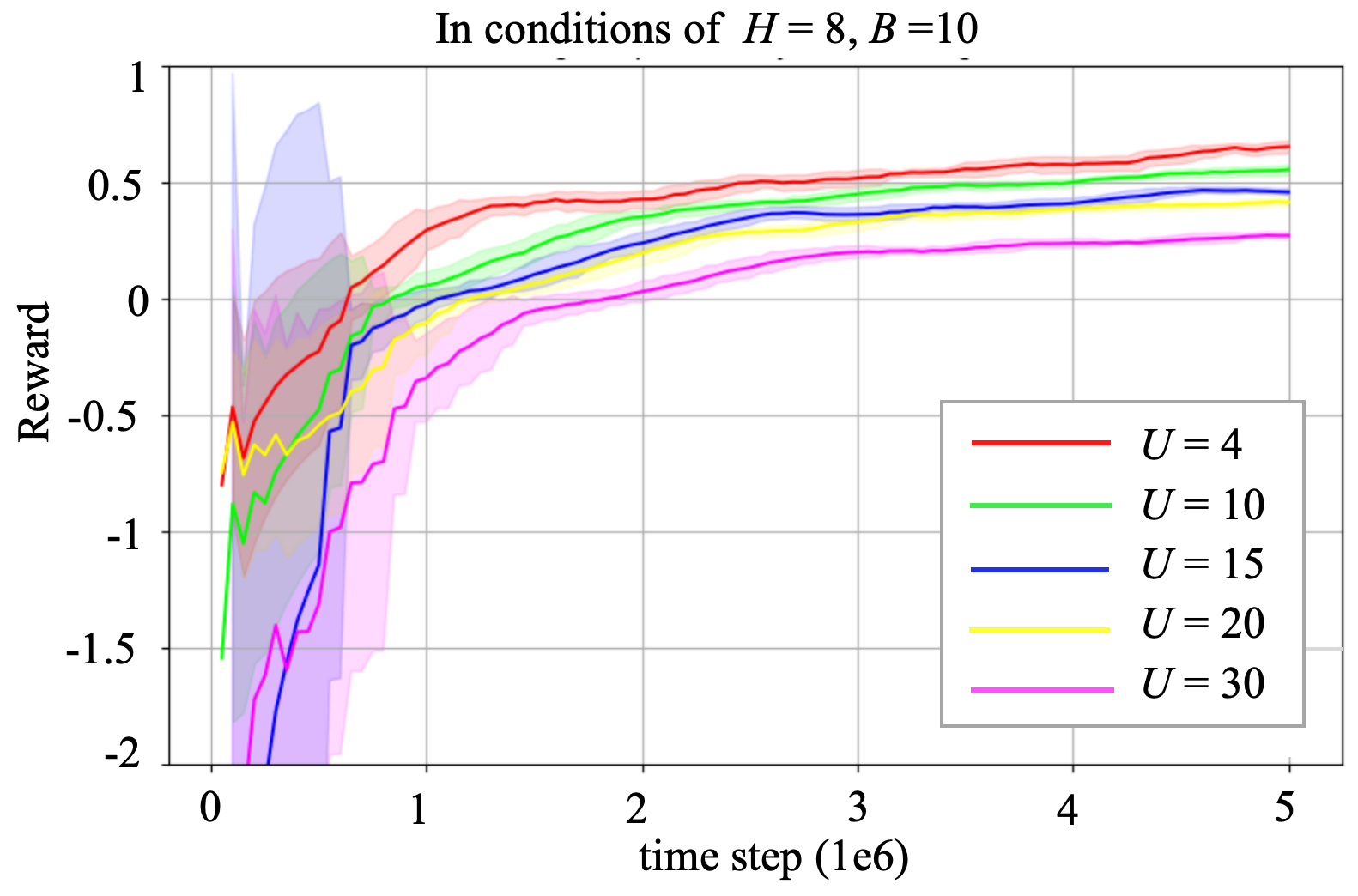} 
\caption{Scalability testing of Pharos based on MAPPO: Training convergence of the average reward per agent for varying scenarios.}
\label{fig:scale_up}
\end{figure}

We identify that computational costs arise mainly during the training and inference phases. The training phase is computation-intensive but a one-time investment, which can be optimized through pre-training. Theoretical analysis of multi-agent collaborative learning indicates that the time complexity at this phase increases approximately linearly with the number of agents. We actually carried out the scalability tests shown in Fig. \ref{fig:scale_up}, which demonstrated the stable convergence of Pharos model training under varying numbers of UAVs. This also aligns with data scales in other leading UAV collaboration studies \cite{huang2024code,yan2023collision,yan2023pascal}.

During the inference phase, each UAV agent executes in a parallel manner and only needs to run a lightweight policy network for forward computing, thereby achieving a low-latency response. Experiments demonstrate that the system can achieve an average inference delay of approximately 100 ms per time step in the current dynamic environment (see the annotations in Fig. \ref{fig:scale_up} and Tab. \ref{tab:human_fear} respectively), making it suitable for the dynamically evolving urban low-altitude environment.

\section{Conclusion and Future Work}
We have developed an airspace management system for multi-UAV collaborative coordination based on the MAPPO algorithm. Through comprehensive tests of collision avoidance, human fear impact, and space utilization, we demonstrated Pharos's effectiveness, accuracy, and performance. The focus of our research in the future is as follows: \textbf{(1) Cross-zone coordination for UAVs}: We will investigate partitioning wide-area airspace and enabling cross-zone UAV coordination to improve system efficiency and scalability. \textbf{(2) Entropy-driven for UAV on-demand coordination}: Based on the spatial entropy mentioned in the experiment evaluation metric, we will apply it to on-demand access and on-demand coordination of UAVs in urban settings. \textbf{(3) 4D spatio-temporal data management}: We will develop a 4D data management engine to handle the complex 3D spatial and temporal dynamics of urban low altitude.
%We will construct a distributed communication architecture based on cloud-edge continuum, providing reliable support for crowdsourced uploading of situational awareness data and stable communication between the Pharos center and UAVs.
\bibliographystyle{ACM-Reference-Format}
\bibliography{sample-base}

\end{document}